\documentclass[12pt]{article}
\usepackage{latexsym,amssymb,amsmath}
\usepackage{epsf}
\usepackage[dvips]{graphicx}
\usepackage{psfrag}
\newcommand{\bee}{\begin{eqnarray}}
\newcommand{\eee}{\end{eqnarray}}
\newcommand{\be}{\begin{equation}}
\newcommand{\ee}{\end{equation}}
\newcommand{\ds}{\displaystyle}
\def\conj#1{\stackrel{*}{#1}}

\begin{document}

\begin{center}
{\bfseries TWO-PHOTON EXCHANGE AND POLARIZATION PHYSICS\\ IN ELECTRON-PROTON SCATTERING}

\vskip 5mm

A.P.~Kobushkin$\dag$ and D.L.~Borisyuk

\vskip 5mm

{\small
{\it Bogolyubov Institute for Theoretical Physics, National Academy of Sciences of Ukraine\\
Metrologicheskaya street 14B, 03680, Kiev, Ukraine
}
\\
$\dag$ {\it E-mail: kobushkin@bitp.kiev.ua}
}
\end{center}

\vskip 5mm

\begin{center}
\begin{minipage}{150mm}
\centerline{\bf Abstract}
We discuss effects of two-photon exchange (TPE) in various observables of the elastic $ep$-scattering. The imaginary part of the TPE amplitude manifests in target and beam normal spin asymmetries. The real part contributes to the cross-section. A model-independent phenomenological analysis of experimental data is performed. Its results are compared with theoretical calculations.
\end{minipage}
\end{center}

\vskip 10mm

%
The experimental study of the electron-nucleon scattering gives important information about the electromagnetic structure of the nucleon. Experimental data for the elastic scattering are usually expressed in terms of two fundamental observables, the electric and magnetic form factors (FFs), $G_E$ and $G_M$, which parametrize the $\gamma NN$ vertex with two on-mass shell nucleons
\bee
 &\Gamma_\mu=\gamma_\mu F_1(q^2)-[\gamma_\mu,\gamma_\nu]\ds{\frac{q^\nu}{4M}}F_2(q^2),\label{1}\\
&G_E(q^2)=F_1(q^2)-\tau F_2(q^2),\qquad G_M(q^2)=F_1(q^2)+ F_2(q^2),\qquad \tau = - q^2/4M^2,\nonumber
\eee
where 
$M$ is the nucleon mass. 

\begin{figure}[t]
\psfrag{k}{$k$}\psfrag{k1}{$k'$}\psfrag{p}{$P$}\psfrag{p1}{$P'$}\psfrag{q}{$q$}
\psfrag{k2}{$k''$} \psfrag{q1}{$q_1$}\psfrag{q2}{$q_2$}
\centerline{
\epsfysize=25mm\epsfbox{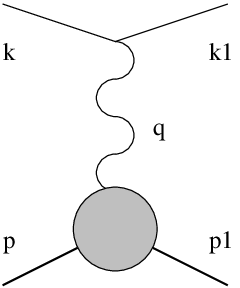}\hspace{3.cm}\epsfysize=25mm\epsfbox{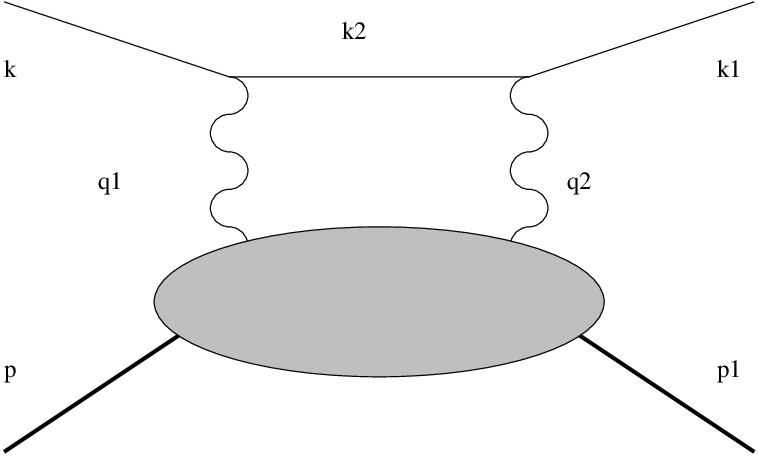}}
 \caption{One-photon exchange (left) and two-photon exchange (right) diagrams.}\label{fig:1}
\end{figure}

The important question is how FFs are extracted from experimental data on $eN$-scattering. There are two procedures, the Rosenbluth (LT separation) method and the polarization transfer (PT) method. Both of them are based on the one-photon exchange (OPE) approximation, see left panel of Fig.~\ref{fig:1}. The LT separation \cite{Rosenbluth} is based on linearity of the reduced cross section $\sigma_R$ in $\varepsilon$:
\be
\label{2}
\sigma_R \equiv \frac{E^2 q^2 (1-\varepsilon)}{2\pi\alpha^2}\frac{d\sigma}{dq^2}=\varepsilon G^2_E(q^2)+\tau G^2_M(q^2),
\qquad \varepsilon=\left[ 1+2(1+\tau)\tan^2\tfrac{\theta_e}2 \right]^{-1}. 
\ee
In Eq.~(\ref{2}) $E$ is electron energy in the lab. frame and $\theta_e$ is the scattering angle in the lab. frame. Measuring $\sigma_R$ at different $\varepsilon$, but fixed $q^2$ one can determine $G_E$ and $G_M$ as square root from the slope and intercept of this plot, respectively.

Basing on the calculations of Akhiezer and Rekalo \cite{AkhiezerRekalo} one can derive that the ratio of the FFs is connected with the ratio of transversed to longitudinal polarizations, $P_t/P_l$, of the final proton in collision of longitudinally polarized electron with unpolarized proton:
\be\label{PT}
\frac{G_E}{G_M} = - \frac{P_t}{P_l}\times
\frac{(E+E')}{2M}\tan \frac{\theta_e}{2} ,
 \qquad \left\{
\begin{array}{l}
\vec {e}+{p}\to {e}+{p}^{\uparrow}\\
\vec {e}+{p}\to {e}+\vec{p}
\end{array}\right.
\ee
($E$ and $E'$ are energies of the incoming and outgoing electrons in the lab. frame).
This allows to measure the ratio $G_E/G_M$ directly.

The main surprise, which we got from the PT measurements, is that such measurements give the ratio $G_E^p/G_M^p$ different from the results of LT separation (for the further references see talks of C.~Perdrisat and V.~Punjabi at this Seminar). This means that higher order perturbative terms have to be included in analysis of the elastic $eN$-scattering.

There are two types of the higher order perturbation effects which should be taken into account
\begin{itemize}
 \item Contribution of soft photon radiation beyond Tsai approximation \cite{Tsai}. Some of such estimates were obtained in \cite{TyonMaximon,Kuraev}.
 \item Effects coming from two-photon exchange (TPE) diagram (right panel of Fig.~\ref{fig:1}).
\end{itemize}

In the present talk we discuss how TPE effects can be extracted from experiment and compared with theoretical calculations.

%
%
TPE amplitude is complex. Interference of its real part with OPE amplitude contributes to the cross-section. We will discuss this point later, but now we will concentrate on the imaginary part. It manifests in one particle normal spin asymmetries
\be
A_n,\  B_n=\frac{\sigma_\uparrow-\sigma_\downarrow}{\sigma_\uparrow+\sigma_\downarrow},
\ee
where $\sigma_\uparrow$ and $\sigma_\downarrow$ are cross sections, when spin of one of the particles is directed up or down to the scattering plane. We are speaking about target asymmetry ($A_n$) or beam asymmetry ($B_n$) depending on whether the incoming proton or electron is polarized.

The imaginary part can be expressed through the unitarity condition
\begin{equation} \label{unit}
T_{fi}-\conj T_{if} = i\sum_n T_{fn} \conj T_{in},
\end{equation}
where $i$ and $f$ are initial and final states, respectively, $n$ is 
intermediate state and $T_{fi}$ are $T$-matrix elements.
In our case we can use one-photon exchange
amplitudes in the right-hand side of (\ref{unit}). Then we obtain
\begin{equation} \label{unit2}
\includegraphics[height=0.1\textheight]{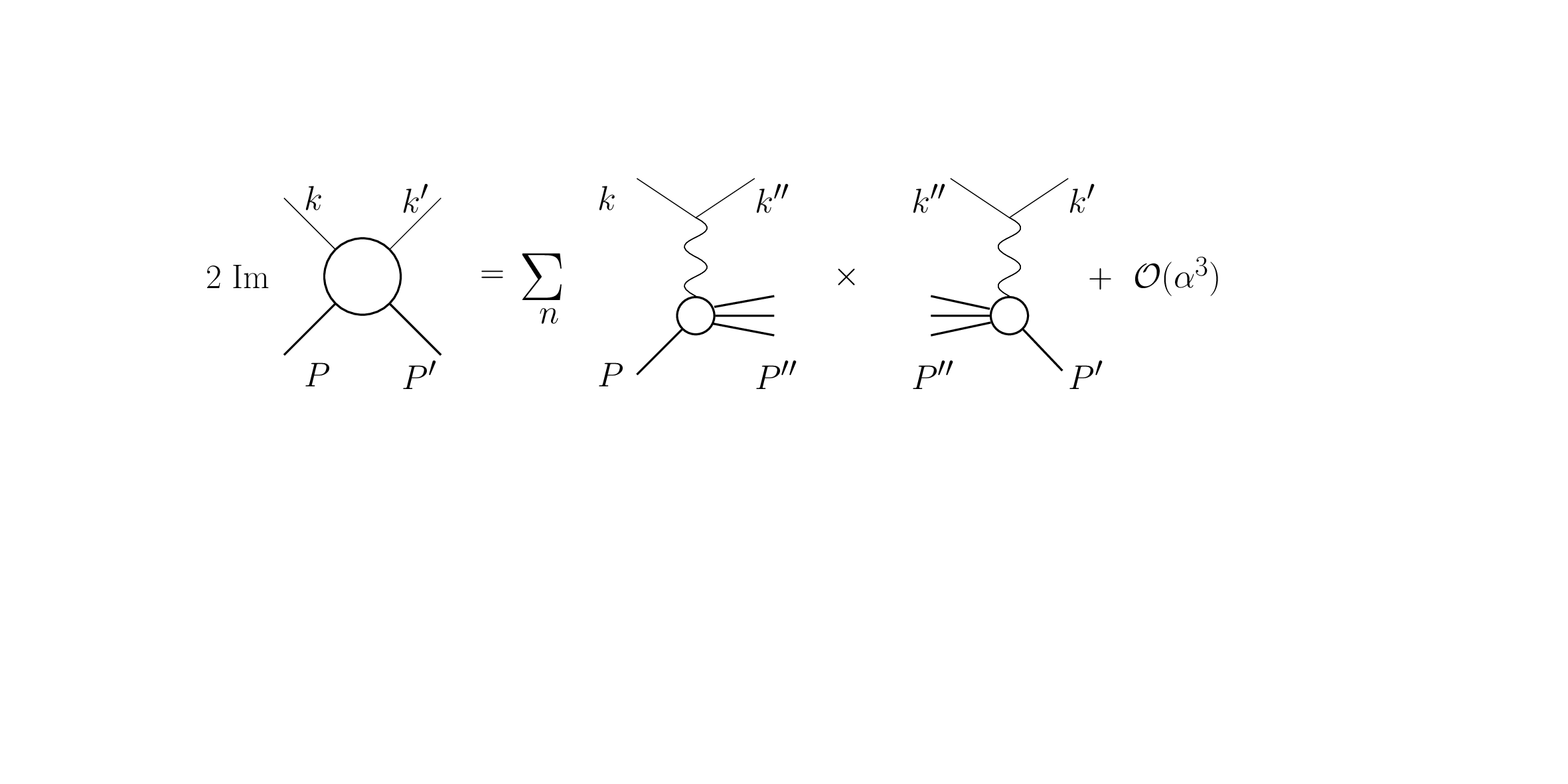} 
\end{equation}
In calculations \cite{bib1} we used the proton (elastic contribution) and the most important resonances from the first, second and third resonance regions, $\Delta(1232)$, $S_{11}(1520)$, $D_{13}(1535)$, $F_{15}(1680)$ and Roper resonance $P_{11}(1440)$ as intermediate states.

\begin{figure}[t]
  \includegraphics[height=0.22\textheight]{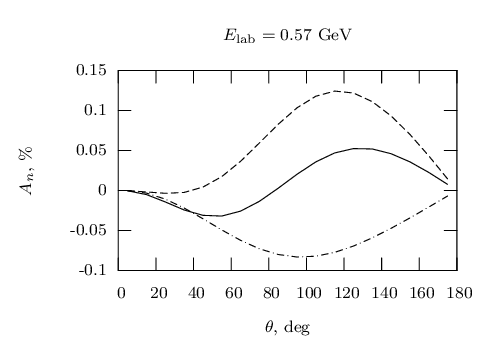}
  \includegraphics[height=0.22\textheight]{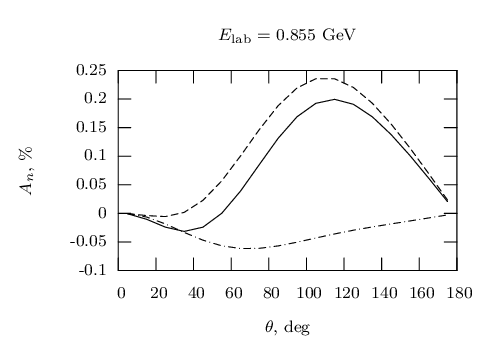}\\
  \includegraphics[height=0.22\textheight]{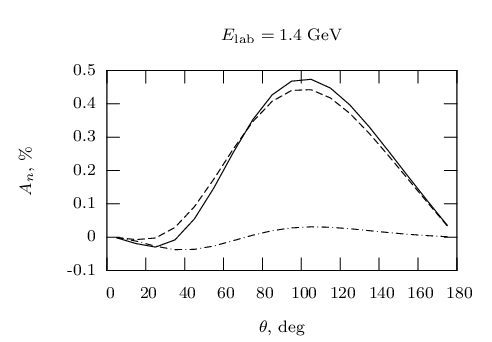}
  \includegraphics[height=0.22\textheight]{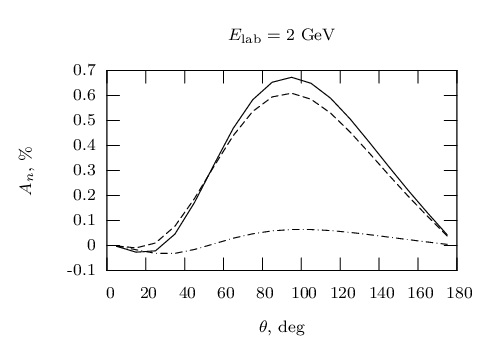}
\caption{\label{fig:2} Target normal spin asymmetry for different electron lab. energies. Elastic contribution (dashed), inelastic contribution (dash-dotted), total (solid).}
\end{figure}

The results for the target asymmetry are given in Fig.~\ref{fig:2}. Note that at high beam energy contributions 
from the resonances tend to cancel each other.  The elastic contribution dominates at low 
energy ($E_\mathrm{lab} < 300 $ MeV) and at $E_\mathrm{lab} > 1300 $ MeV. 
At high energy it is nontrivial result, which has interesting consequences. Since the real and imaginary parts are connected
(via the dispersion relations), we may expect that the real part will also be defined mostly by the proton contribution.

\begin{figure}
  \includegraphics[height=0.18\textheight]{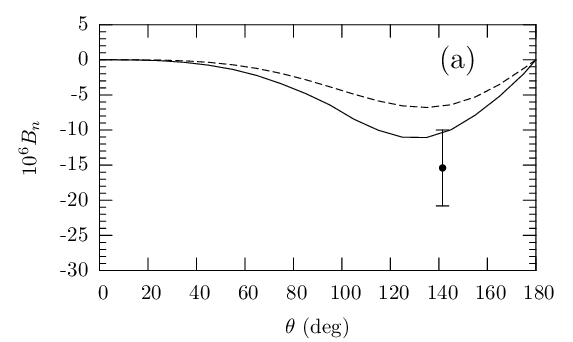}
  \includegraphics[height=0.18\textheight]{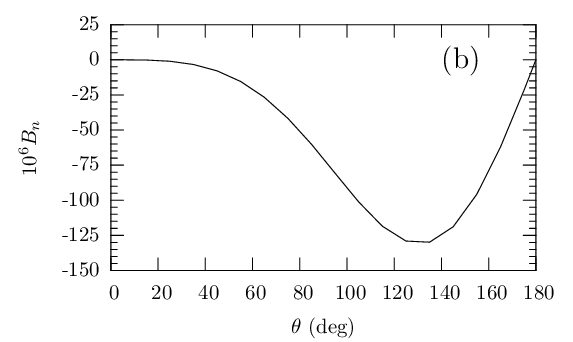}\\
  \includegraphics[height=0.18\textheight]{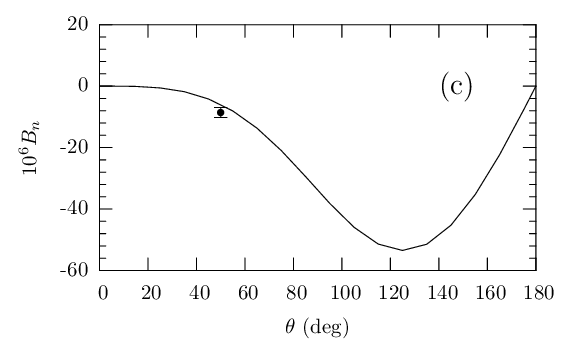}
  \includegraphics[height=0.18\textheight]{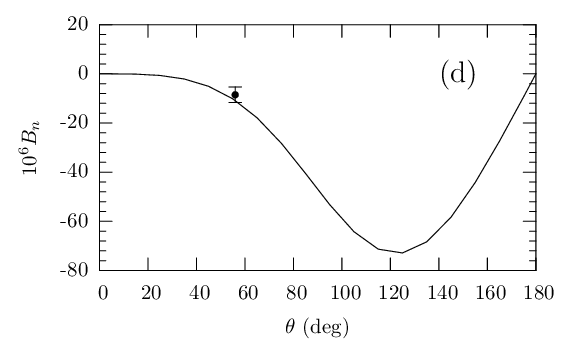}
 \caption{\label{fig:3} The beam normal spin asymmetry of the elastic $ep$ scattering at electron energies (a) 0.2 GeV, dashed line - without elastic contribution, (b) 0.3 GeV, (c) 0.57 GeV and (d) 0.855 GeV. Experimental points are from \cite{SAMPLE}.} 
\end{figure}

A theoretical analysis of $B_n$ was done within different approaches \cite{AfanAkushMer,DiakR-M,GGV,Pasq,AfanMer,Gorchtein,Beam}. The asymmetry $B_n$ is proportional to the fine structure constant $\alpha\approx1/137$, as well as to the electron mass $m$. An important feature is that the asymmetry contains terms, proportional to the large logarithms, $\ln^2(Q^2/m^2)$ and $\ln(Q^2/m^2)$
\be 
B_n=m\left(A \ln^2\frac{Q^2}{m^2}+B\ln\frac{Q^2}{m^2}+C\right).
\ee
In Ref.~\cite{Beam} the general expression for $B_n$ was obtained in the leading logarithm approximation (term proportional to $\ln^2({Q^2}/{m^2})$). This result is valid for a wide range of scattering angles, besides very forward kinematics. When $Q^2\to 0$ the leading logarithm term behaves as 
\begin{equation} \label{asymp}
B_n^{(2)} \sim Q^3 \ln^2(Q^2/m^2).
\end{equation}
Alternatively,  in the limit of forward scattering the terms with the lowest power of $Q$ are \cite{AfanMer}
\begin{equation} \label{AfBn}
B_n \approx B_n^{(1)} \sim Q \ln (Q^2/m^2).
\end{equation}
%
In general, the distinct contributions (\ref{asymp}) and (\ref{AfBn}) should be added together. Comparing them one can derive condition for the validity of leading logarithm approximation
\be \label{app}
 \sin^2 \frac{\theta}{2} \, \ln \frac{Q^2}{m^2} \gg 1.
\ee
The asymmetry $B_n$ calculated in the leading logarithm approximation with three lightest resonances, $P_{33}(1232)$, $D_{13}(1520)$ and $S_{11}(1535)$ as intermediate states is shown in Fig.~\ref{fig:3}. We also add the contribution from the threshold pion production in the $s$-wave. It is especially important for $E_{\rm lab} = 0.2$ GeV.
%
%
%

The role of TPE effects in explanation of the discrepancy between results for the ratio $G_E^p/G_M^p$ measured with LT and PT techniques has been discussed by many authors, both phenomenologically \cite{GV,Tvaskis,bib3,C} and theoretically \cite{Blunden,BMTdelta,GPD,we,bib2}.
Neglecting the electron mass the general expression for the elastic $eN$-scattering amplitude reads \cite{GV}
\begin{equation}
 {\cal M} = \frac{4\pi\alpha}{Q^2} \bar u'\gamma_\mu u \cdot
 \bar U' \left(\tilde F_1 \gamma^\mu - \tilde F_2 
 [\gamma^\mu,\gamma^\nu] \frac{q_\nu}{4M} +
 \tilde F_3  k_\nu \gamma^\nu \frac{P^\mu}{M^2}
 \right) U.
\end{equation}
Invariant amplitudes (also called generalized FFs) $\tilde F_i$
are complex scalar functions of two kinematic variables, say, $Q^2$ and $\varepsilon$. In the framework of OPE $\tilde F_1$ and $\tilde F_2$ are reduced to Dirac and Pauli FFs and $\tilde F_3$ vanishes. 

For our purpose it is convenient to introduce linear combinations \cite{bib3}
\begin{equation} \label{calG}
 \mathcal G_E = \tilde F_1 - \tau \tilde F_2 + \frac{\nu}{4M^2} \tilde F_3, \qquad
 \mathcal G_M = \tilde F_1 +\tilde F_2 + \varepsilon \frac{\nu}{4M^2} \tilde F_3,\qquad
 \mathcal G_3 = \frac{\nu}{4M^2}\tilde F_3,
\end{equation}
where $\nu=(k+k')(P+P')$ and the reduced cross section for unpolarized particles reads
\begin{equation}\label{CS_unpol}
 \sigma_R = \varepsilon |\mathcal G_E|^2 + \tau |\mathcal G_M|^2 + \tau\varepsilon^2\frac{1-\varepsilon}{1+\varepsilon}|\mathcal G_3|^2.
\end{equation}
Dropping terms proportional to $\alpha^2$ Eq.~(\ref{CS_unpol}) can be written like the Rosenbluth formula
\begin{equation}\label{CS_unpol1}
 \sigma_R = \varepsilon \mathcal G_E^2 + \tau \mathcal G_M^2 + O(\alpha^2),
\end{equation}
but LT separation of FFs $\mathcal G_E$ and $\mathcal G_M$ cannot be done, because  now the FFs are functions of two variables (in the last equation and further the real parts of amplitudes are understood).
The amplitudes can be decomposed as
\begin{equation}\label{Expansion}
 \mathcal G_E(Q^2,\varepsilon) = G_E(Q^2) + \delta G_E^{(T)} (Q^2,\varepsilon) + \delta\mathcal G_E(Q^2,\varepsilon) 
\end{equation}
and similarly for $\mathcal G_M$. Here $\delta G_{E,M}^{(T)}$ and $\delta\mathcal G_{E,M}$ are TPE corrections of order $\alpha$, where $\delta G_{E,M}^{(T)}$ denotes the correction, calculated by Tsai \cite{Tsai}. All infrared divergence is contained in it. With Eq.~(\ref{Expansion}) one gets at
\begin{equation}
 \sigma_R = \varepsilon G_E^2 + \tau G_M^2 + 2 \varepsilon G_E \delta\mathcal G_E + 2 \tau G_M \delta\mathcal G_M + 2 \varepsilon G_E \delta G_E^{(T)} + 2 \tau G_M \delta G_M^{(T)}.
\end{equation}
The terms with $\delta G_{E,M}^{(T)}$  are usually subtracted from data by experimenters as a part of radiative corrections, so the cross sections, which are used as an input for the LT separation, are
\begin{equation} \label{Rosen}
 \sigma_R = \varepsilon G_E^2 + \tau G_M^2 + 2 \varepsilon G_E \delta\mathcal G_E + 2 \tau G_M \delta\mathcal G_M.
\end{equation}
The key point is that when  $\tau \gtrsim 1$ the contribution of $G_M$ is enhanced with respect to $G_E$ by the proton magnetic moment $\mu$, which gives a factor of 3 (and even more according to PT data). So we have
\begin{equation}
 \tau G_M^2 \gg \varepsilon G_E^2 \gg 2 \varepsilon G_E \delta\mathcal G_E, \ \ \tau G_M^2 \gg 2 \tau G_M \delta\mathcal G_M \gg 2 \varepsilon G_E \delta\mathcal G_E.
\end{equation}
Therefore the term $2 \varepsilon G_E \delta\mathcal G_E$ can be safely neglected. Instead, because $2 \tau G_M \delta\mathcal G_M$ can be comparable with the term $\varepsilon G_E^2$ it should strongly affect the results of LT separation. 
The analyses \cite{Tvaskis,C} show that $\sigma_R$ is linear function of $\varepsilon$, which means that $\delta \mathcal G_M$ is also approximately linear
\begin{equation} \label{linear}
 \delta \mathcal G_M(Q^2,\varepsilon) = [a(Q^2) + \varepsilon b(Q^2)] G_M(Q^2).
\end{equation}
The coefficient $a$ yields only a small contribution to the large $\varepsilon$-independent term in $\sigma_R$ and does not change $\varepsilon$-dependent term, thus we also may neglect it.
After that
\begin{equation}
 \sigma_R = \tau G_M^2 + \varepsilon (G_E^2 + 2 \tau b \, G_M^2 )
\end{equation}
and the FF ratio squared obtained by the LT method is actually
\begin{equation} \label{R_LT}
\left. \left({G_E}/{G_M} \right)^2 \right|_{LT} \equiv R_{LT}^2 = {G_E^2}/{G_M^2} + 2 \tau b.
\end{equation}
On the other hand, the FF ratio obtained in the PT experiments (even with the inclusion of TPE)
is close to true FF ratio 
%
%
%
\begin{equation} \label{R_P}
 R_{PT} \approx {G_E}/{G_M}.
\end{equation}
%
Finally the TPE correction slope can be written as
\begin{equation}
 b = \left(R_{LT}^2 - R_{PT}^2\right) / 2\tau.
\end{equation}
Fitting data \cite{PT,LT} with polynomials in $Q^2$ of the third order one gets:
%
$\ b = - 0.0101/Q^2 + 0.0314 + 0.0008 Q^2 + 0.0005 Q^4$.
%
The coefficient $a$ cannot be derived from data.
\begin{figure}[t]
\centering
\includegraphics[width=0.45\textwidth]{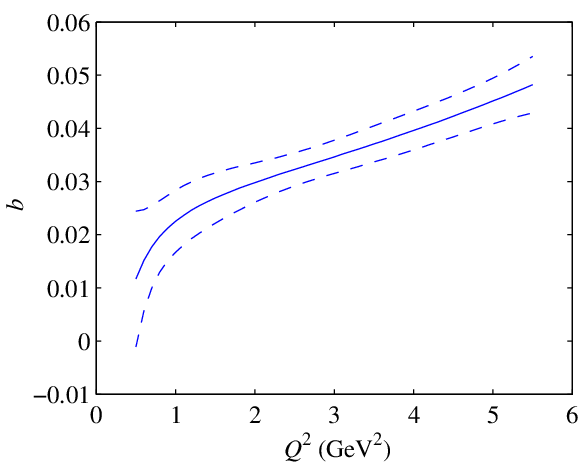}\hfill
\includegraphics[width=0.45\textwidth]{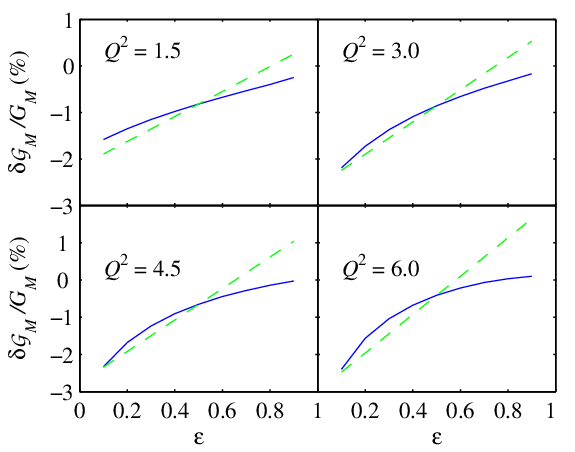}
\caption{Left panel: Extracted TPE correction slope $b(Q^2)$; the dashed curves indicate estimated errors.
Right panel: Comparison of extracted (dashed lines) and calculated (solid curves) in Refs.\cite{we,bib2} values of TPE amplitude $\delta\mathcal G_M/G_M$.}\label{coefb}
\end{figure}

We display the $Q^2$ dependence of $b$ in Fig.~\ref{coefb}. It varies between 0.01 and 0.05; note that the relative size of TPE correction $\delta \mathcal G_M/G_M = a + \varepsilon b$ will never exceed 0.025 if the unknown coefficient $a$ is negative and equals about $-b/2$. This agrees with the assumptions made at the very beginning of our phenomenological analysis of LT and PT data.

In Fig.~\ref{coefb} we also plot the elastic part of $\delta \mathcal G_M$, calculated as described in Ref.~\cite{we,bib2} and straight line according to Eq.(\ref{linear}) with $b$ given above and $a$ chosen arbitrarily. A qualitative agreement is seen at all values of $Q^2$. The gap between the curve and the line is small and should be possibly attributed to inelastic contribution, which was not taken into account in theoretical calculations \cite{we,bib2}.
Thus inclusion of TPE is enough to explain the disagreement between the results of Rosenbluth and PT methods.
%
%

\end{document}